\documentclass[letterpaper,showpacs,superscriptaddress,twocolumn,aps,pra]{revtex4-1}

\usepackage{graphicx}
\usepackage{amssymb}

\begin{document}

\title{High-threshold topological quantum error correction against biased noise}

\author{Ashley M. Stephens}\email{astephens@nii.ac.jp}
\affiliation{National Institute of Informatics, 2-1-2 Hitotsubashi, Chiyoda-ku, Tokyo 101-8430, Japan}

\author{William J. Munro}
\affiliation{NTT Basic Research Laboratories, 3-1 Morinosato-Wakamiya, Atsugi, Kanagawa 243-0198, Japan}
\affiliation{National Institute of Informatics, 2-1-2 Hitotsubashi, Chiyoda-ku, Tokyo 101-8430, Japan}

\author{Kae Nemoto}
\affiliation{National Institute of Informatics, 2-1-2 Hitotsubashi, Chiyoda-ku, Tokyo 101-8430, Japan}

\date{\today}

\begin{abstract}
Quantum information can be protected from decoherence and other errors, but only if these errors are sufficiently rare. For quantum computation to become a scalable technology, practical schemes for quantum error correction that can tolerate realistically high error rates will be necessary. In some physical systems, errors may exhibit a characteristic structure that can be carefully exploited to improve the efficacy of error correction. Here, we describe a scheme for topological quantum error correction to protect quantum information from a dephasing-biased error model, where we combine a repetition code with a topological cluster state. We find that the scheme tolerates error rates of up to 1.37\%--1.83\% per gate, requiring only short-range interactions in a two-dimensional array.
\end{abstract}

\maketitle

Many physical systems have been identified as candidates for the qubits in a quantum computer \cite{Ladd2010}, and each of these systems will suffer from noise with distinct structure. Quantum error correction \cite{Gottesman1} can suppress a remarkably wide range of noise, including long-range correlated noise, Gaussian noise, and qubit loss \cite{Aharonov06,Aliferis06,Ng08,Barrett1}. One particularly interesting noise model is so-called biased noise, where the characteristic time for dephasing (loss of phase coherence) is much shorter than the equivalent time for population relaxation (exchange of energy with the environment) \cite{Gourlay1,Evans1,Stephens1}. This may be the case in several of the most promising physical systems, including qubits based on superconducting circuits, semiconductor spins, trapped ions, and negatively charged nitrogen vacancy centers in diamond \cite{Taylor05,Astafiev1,Brito08,Martinis02,Ospelkaus08,Doherty13}. Aliferis and Preskill among others have proposed schemes for quantum error correction using a restricted set of quantum gates for which biased noise is a reasonable assumption \cite{Aliferis3,Aliferis4,Brooks1,Brooks2}. In particular, the fundamental physical operation in these schemes is the two-qubit controlled-phase gate---derived from the two-qubit Ising Hamiltonian---which commutes with noise due to dephasing, thereby preserving the bias. These schemes indicate that biased noise can be successfully exploited. Their drawback is that they require long-range interactions to achieve arbitrarily low logical error rates, which may be difficult to achieve with sufficient fidelity.

A promising alternative may involve topological quantum error correction \cite{Kitaev2003,Dennis2002, Raussendorf3,Fowler1}. Assuming that only short-range interactions are available, some topological codes tolerate error rates more than two orders of magnitude higher than concatenated codes \cite{Svore07, Stephens08,Stephens09,Spedalieri09,Raussendorf2007,Fowler1,Wang10,Wang11}. However, existing schemes for topological error correction do not exploit biased noise, leaving room for significant improvement \cite{Loss12}. In this article, we describe a scheme for topological error correction against biased noise, where a repetition code is used to suppress errors due to dephasing and a topological code is used to suppress the errors that remain. For highly biased noise, we find that the scheme tolerates error rates of up to 1.37\%--1.83\% per gate. In practice, the scheme may operate with error rates approaching one percent per gate, at which other schemes cannot effectively suppress errors. By reducing the requirements for scalable quantum computing in systems where dephasing is the dominant error, our scheme illustrates the importance of tailoring quantum error correction to the wide variety of systems and architectures under consideration.

\begin{figure*}
\begin{center}
\includegraphics[scale=0.50]{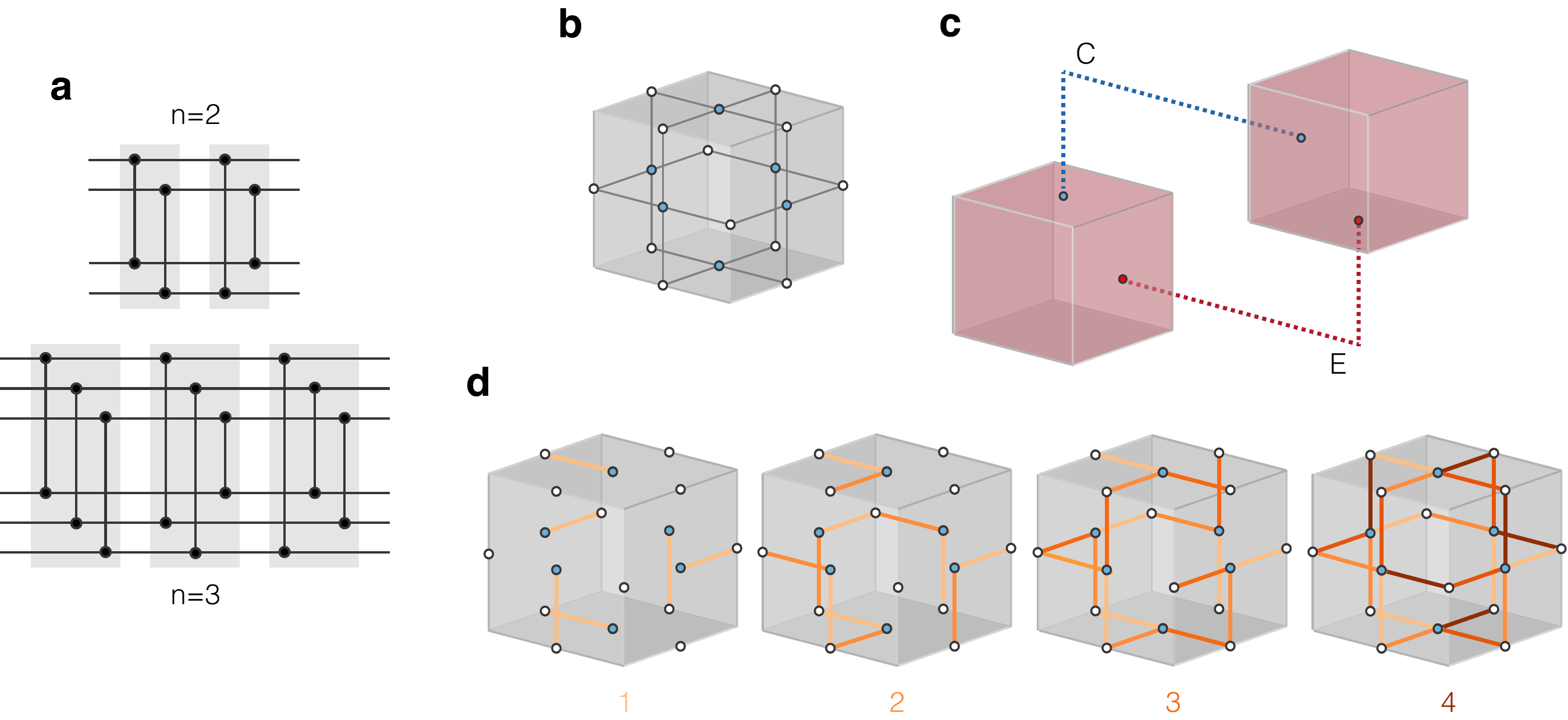}
\end{center}
\vspace{-18pt}
\caption{(Color online) Generating the topological cluster state. (a) Set of physical controlled-$Z$ gates to execute the encoded controlled-$Z$ gate between $\mathcal{C}_{\rm 1}$ code blocks for various values of the repetition code length, $n$. The gates in each grey block can be performed in parallel in one time step, so that no qubit is ever idle in between gates. (b) Elementary cell of the topological cluster state, with qubits on the faces and edges of a three-dimensional lattice. The generators of the stabilizer group of the topological cluster state are $X_i \otimes_{j\in N(q_i)}Z_j\forall i$, where $q_i$ is a qubit and $N(q_i)$ is the set of its four nearest neighbours. The qubits on the six faces of the cell are measured in the $X$ basis to determine the parity of a cell. (c) Chains of $Z$ errors, such as $E$, are revealed by cells with odd-parity at the endpoints, which combine to give an error syndrome. In this case, $C$ is a candidate correction consistent with the syndrome. (d) Order of encoded controlled-$Z$ gates to prepare the topological cluster state from encoded $\mathcal{C}_{\rm 1}$ qubits in the state $\vert\bar+\rangle$.} 
\label{Figure1}
\end{figure*}

{\it Repetition code in the dual basis---} Our scheme is based on a concatenation of two stabilizer quantum codes \cite{Gottesman97}. Following Aliferis and Preskill, the base-level code in our scheme is a length-$n$ quantum repetition code in the dual basis, denoted by $\mathcal{C}_{\rm 1}$ \cite{Aliferis3}. The generators of the stabilizer group of $\mathcal{C}_{\rm 1}$ are 
\begin{equation}
I^{\otimes i} \otimes X_{i+1} \otimes X_{i+2} \otimes I^{\otimes n-i-2}, ~i=\{0,1,\ldots,n-2\}, 
\end{equation}
and the encoded Pauli operators are 
\begin{equation}
\bar{X}=X_1 \otimes I^{\otimes n-1},~\bar{Z}=Z_1 \otimes Z_2 \otimes \ldots \otimes Z_n,
\end{equation}
where $X$ and $Z$ are the usual single-qubit Pauli operators. Because $\vert\bar{+}\rangle=\vert+\rangle^{\otimes n}$, preparation of $\vert\bar{+}\rangle$ involves preparing $\vert+\rangle$ for all $n$ qubits. Similarly, measurement of $\bar{X}$ involves measurement of $X$ for all $n$ qubits and taking a majority vote of the $n$ outcomes. The last basic ingredient is the encoded controlled-phase gate (or, controlled-$Z$) between two $\mathcal{C}_{\rm 1}$ code blocks, which involves $n^2$ physical controlled-$Z$ gates, as shown in Fig.~\ref{Figure1}(a). As $Z$ errors are not spread among the qubits in a code block (they commute with the controlled-$Z$ gates), $\mathcal{C}_{\rm 1}$ protects against $\lfloor (n-1)/2\rfloor$ $Z$ errors per block but offers no protection against $X$ errors.

{\it Topological cluster-state error correction---} The top-level code in our scheme is the topological code associated with a three-dimensional topological cluster state, denoted by $\mathcal{C}_{\rm 2}$ \cite{Raussendorf3}. The elementary cell of the topological cluster state is shown in Fig.~\ref{Figure1}(b). The topological cluster state is divided into three distinct regions: $V$, $D$, and $S$. To enact computation, qubits in $V$ are measured in the $X$ basis,  qubits in $D$ are measured in the $Z$ basis (or are simply absent from the cluster state), and qubits in $S$ are measured in either the $Y$ or $(X+Y)/\sqrt{2}$ basis \cite{Raussendorf3}. Measurements in $D$ are used to define tubular structures known as defects, which encode logical qubits. Defects are braided with each other to enact a set of Clifford gates. Measurements in $S$ are used to prepare magic states \cite{Bravyi05}, which can be distilled using Clifford gates, enabling a universal set of logical gates. For further details, the reader is referred to \cite{Raussendorf3}. Here, we will focus on error correction in $V$. Error correction occurs on the primal lattice and its dual. However, for simplicity, and without loss of generality, we will consider error correction on only the primal lattice. 

After qubits in $V$ are measured in the $X$ basis, error correction in $\mathcal{C}_{\rm 2}$ proceeds by computing the parity of elementary cells---the parity of a cell is equal to the product of the measurement results of the qubits on its six faces. In the absence of errors, the parity of each cell is $+1$. The endpoints of chains of $Z$ errors are revealed by cells with parity $-1$ [see Fig.~\ref{Figure1}(c)], called the syndrome. Error correction involves identifying a set of errors consistent with the syndrome using an appropriate algorithm \cite{Edmonds1}, then applying the corresponding correction \cite{Dennis2002,Fowler1}. Logical errors occur when errors and corrections combine to connect or encircle defects in $D$. This can be made less likely by increasing the separation and circumference of defects, parameterized by distance $d$. For a standard (unbiased) error model, the threshold of topological cluster-state error correction is approximately $6.3 \times 10^{-3}$ per gate \cite{Barrett1}. For error rates lower than this value, increasing $d$ will always reduce the logical error rate.

\begin{figure*}
\begin{center}
\includegraphics[scale=0.50]{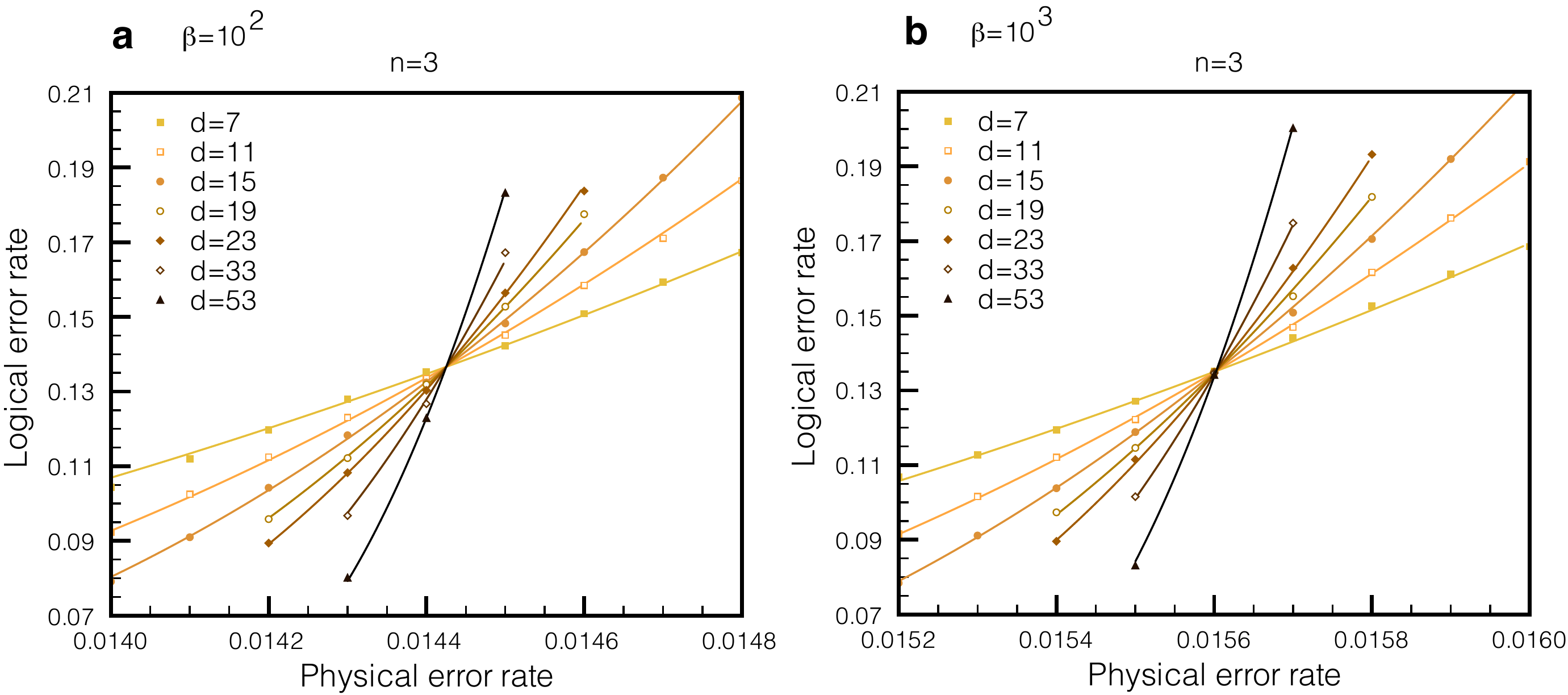}
\end{center}
\vspace{-17pt}
\caption{(Color online) Logical error rate of the concatenated scheme as a function of the physical error rate, $p$, for various values of the topological code distance, $d$, where $n=3$. The curves are best fits accounting for finite-size effects \cite{Wang03}. The threshold is the value of $p$ for which the logical error rate is independent of $d$. For physical error rates below the threshold, increasing $d$ will always result in a lower logical error rate. (a) For $\beta=10^2$, the threshold is $1.443\pm0.001\times 10^{-2}$. The corresponding threshold for $n=2$ is $1.370\pm0.003\times 10^{-2}$. (b) For $\beta=10^3$, the threshold is $1.560\pm0.001\times 10^{-2}$. The corresponding threshold for $n=2$ is $1.421\pm0.003\times 10^{-2}$. Recall that, in our concatenated scheme, error correction in $C_2$ is assisted by information about the reliability of the $C_1$ code blocks. Even the smallest non-trivial repetition code ($n=2$) is useful, as it can detect and locate encoded errors. The threshold for located errors in $C_2$ is determined by the bond-percolation threshold in three dimensions \cite{Barrett1}, which is equal to $24.9\%$ \cite{Lorenz1}, indicating that located errors are much easier to correct than than un-located errors. This fact, combined with the fact that encoded controlled-$Z$ gate requires fewer physical gates for $n=2$ than for $n=3$, may explain why the thresholds for $n=2$ approach those for $n=3$.}
\label{Figure2}
\end{figure*}

{\it Concatenated scheme---} We arrive at our scheme by concatenating $\mathcal{C}_{\rm 1}$ with $\mathcal{C}_{\rm 2}$---that is, the topological cluster state is prepared from qubits encoded in a repetition code. Specifically, the topological cluster state is prepared by preparing encoded qubits in the state $\vert\bar{+}\rangle$ and then applying encoded controlled-$Z$ gates in the order indicated in Fig.~\ref{Figure1}(d). Error correction of $\mathcal{C}_{\rm 1}$ code blocks is performed only at measurement by majority voting, not between the encoded controlled-$Z$ gates. Errors can spread between nearby $\mathcal{C}_{\rm 1}$ code blocks, but the local nature of the circuit to prepare the topological cluster state ensures that errors do not spread beyond a small neighbourhood, regardless of the distance of $\mathcal{C}_{\rm 2}$. This leaves us with $\mathcal{C}_{\rm 1}$ code blocks with some encoded error rate. Then, error correction in $\mathcal{C}_{\rm 2}$ proceeds in the usual way. Topological cluster-state error correction can be mapped to the random plaquette $\mathbb{Z}_2$-gauge model in three dimensions \cite{Dennis2002}, which can tolerate noise of approximately $2.9 \times 10^{-2}$ per qubit \cite{Wang03,Ohno04}. Therefore, $\mathcal{C}_{\rm 2}$ will be effective if the encoded error rate is below this value. The optimal length of the repetition code will minimize the encoded error rate and will be a function of the physical error rate and the bias. Lastly, whenever a $\mathcal{C}_{\rm 1}$ code block is measured, the conditional probability of an encoded error is approximated and used to more accurately identify residual errors in $\mathcal{C}_{\rm 2}$ \cite{Poulin1,Aliferis3,Evans2}.

{\it Monte Carlo simulations---} To estimate the performance of our scheme at high physical error rates, we perform Monte Carlo simulations of topological cluster states. The set of physical operations required in our concatenated scheme is preparation of the state $\vert+\rangle=(\vert0\rangle+\vert1\rangle)/\sqrt{2}$, measurement of $X$, and the controlled-$Z$ gate. We simulate all circuits under a stochastic error model parameterized by the physical error rate, $p$, and the dephasing bias, $\beta$. Our error model is motivated by physical systems in which the computational basis states correspond to the energy eigenstates of an unperturbed qubit. In this case, we may expect that noise due to fluctuations in energy levels---which will be manifested as dephasing---will be much stronger than noise due to transitions between energy eigenstates \cite{Aliferis3,Aliferis4}. Specifically, in our error model, erroneous controlled-$Z$ gates are modelled by perfect controlled-$Z$ gates followed by dephasing-biased noise, where we treat $X$ and $Z$ errors independently---with probability $p/\beta$ an error is chosen randomly from the set $\{I\otimes X,X\otimes I,X\otimes X\}$ and with probability $p$ an error is chosen randomly from the set $\{I\otimes Z,Z\otimes I,Z\otimes Z\}$. Erroneous state preparation occurs with probability $p$ and is modelled by perfect state preparation followed by a Pauli error chosen randomly from the set $\{X,Y,Z\}$. Similarly, erroneous measurement occurs with probability $p$ and is modelled by perfect measurement preceded by a Pauli error chosen randomly from the set $\{X,Y,Z\}$..

In our simulations, we keep track of Pauli errors as they propagate through the full concatenated circuit as the topological cluster state is prepared from encoded $\mathcal{C}_{\rm 1}$ qubits. For each instance of errors, upon measurement of the topological cluster state, the error syndrome is calculated and converted to a weighted graph, where cells with parity $-1$ are joined by edges with weight related to the distance between them. We infer a correction operation by performing minimum-weight matching on the graph. This is done with the Blossom V implementation of Edmonds' minimum-weight perfect matching algorithm \cite{Edmonds1,Kolmogorov1}. Error correction fails if the initial errors combined with any corrections form a logical operator. Logical error rates for fixed parameters are averaged over no fewer than $10^5$ independent trials and, for simplicity, we assume periodic boundary conditions. Thresholds are calculated by fitting data to a universal scaling function, following Wang {\it et al.~}\cite{Wang03}.

\begin{figure}
\begin{center}
\includegraphics[scale=0.50]{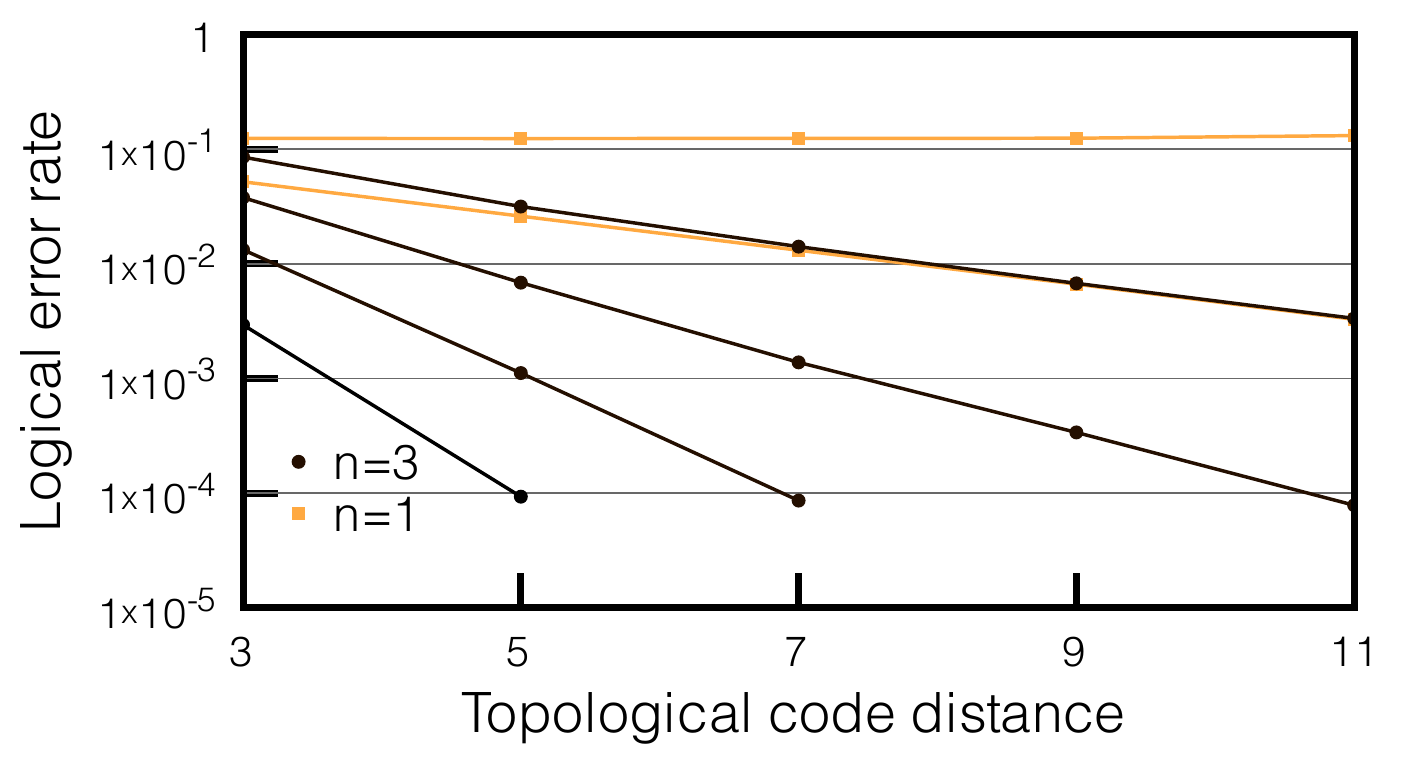}
\end{center}
\vspace{-17pt}
\caption{(Color online) Logical error rate with ($n=3$) and without ($n=1$) the underlying repetition code as a function of the topological code distance, $d$, for $\beta=10^3$ and $p=\{1.25,1.00,0.75,0.50\} \times 10^{-2}$ ($n=3$) and $p=\{0.75,0.50\} \times 10^{-2}$ ($n=1$), in order from top to bottom. All cases are below their respective thresholds, except the $n=1$, $p=0.75\times10^{-2}$ case, which is marginally above the threshold of $\sim0.742\times10^{-2}$. We note that the corresponding curves for $n=2$ (not shown) are qualitatively the same as for $n=3$, as these cases are both below their respective thresholds in this regime.}
\label{Figure3}
\end{figure}

We are interested in the threshold error rate of the concatenated scheme, below which the logical error rate can be suppressed arbitrarily by increasing the distance of $\mathcal{C}_{\rm 2}$. The threshold is a function of the bias, $\beta$, and the length of the repetition code, $n$. In the case where $n=3$ and $\beta=10^2$ ($\beta=10^3$), we find that the threshold of the concatenated scheme is $1.443\pm0.001\times 10^{-2}$ ($1.560\pm0.001\times 10^{-2}$), as shown in Fig.~\ref{Figure2}. If the bias is greater, or if the physical error rate is lower, a larger repetition code may be more effective. These thresholds are significantly higher than the threshold of topological cluster-state error correction without the underlying repetition code ($n=1$), which, for an identical dephasing-biased error model ($\beta=10^3$), we find to be $7.423\pm0.001\times 10^{-3}$. This indicates that there is a range of physical error rates for which suppression of the logical error rate is weak or non-existent without the underlying repetition code, as shown in Fig.~\ref{Figure3}.

We also consider the case where the three-dimensional arrangement of qubits is projected to a two-dimensional plane, noting that it is sufficient to prepare only two adjacent slices of the cluster at a time with a fixed number of qubits \cite{Raussendorf2007}. In this case, initialization and measurement may be performed using the same (nondestructive) physical operation, decreasing the number of error prone operations in the circuit. In the case where $n=3$ and $\beta=10^3$, we find that the threshold is increased to $1.693\pm0.001\times 10^{-2}$. If the probability of a measurement error is reduced to $p/100$, then the threshold is again increased to $1.830\pm0.001\times 10^{-2}$. Further improvement in the threshold may be found by considering a more sophisticated algorithm for interpreting the combined syndrome of $\mathcal{C}_{\rm 1}$ and $\mathcal{C}_{\rm 2}$. On the other hand, alternative decoding algorithms may have a lower threshold but may be more practical for large codes due to their reduced complexity \cite{Duclos-Cianci2,Bravyi1}.

{\it Discussion---} In summary, we have found that the threshold for topological quantum error correction can be significantly increased by exploiting biased noise, without compromising the local nature of the scheme. The cost is a small constant increase in the overhead and more extensive (yet still local) interactions between nearby qubits. This tradeoff is a simple one, but a full assessment of any scheme for fault-tolerant quantum computing will involve a number of architectural considerations and depend on a range of factors \cite{Devitt}. As such, it will be instructive to develop and study specific architectures to implement our scheme. As suggested by Aliferis {\it et al.}, the existence of effective schemes for quantum error correction against biased noise has implications for the design of qubits and gates \cite{Aliferis4}. For example, eliminating the need to implement Hadamard gates at the physical level may allow for simpler and more reliable implementations of other physical operations. Lastly, as the efficacy of our scheme depends on the strength of the bias, it will be useful to estimate physically realistic values of $\beta$ for various architectures. This has been done for an architecture based on superconducting circuits \cite{Aliferis4}, but not in other physical systems. This is the subject of further work.

{\it Acknowledgements---} The authors acknowledge financial support from the FIRST Project and NICT in Japan.

\bibliographystyle{unsrt}

\end{document}